# Emergence of Winner-takes-all Connectivity Paths in Random Nanowire Networks

Hugh G. Manning[1,3,†], Fabio Niosi[1,3,†] Claudia Gomes da Rocha[2,3,†], Allen T. Bellew[1,3], Colin O'Callaghan[2,3], Subhajit Biswas[4], Patrick Flowers[5], Ben J. Wiley[5], Justin D. Holmes[4], Mauro S. Ferreira[2,3], and John J. Boland[1,3*]

[1] School of Chemistry, Trinity College Dublin, Dublin 2, Ireland.

[2] School of Physics, Trinity College Dublin, Dublin 2, Ireland.

[3] Centre for Research on Adaptive Nanostructures and Nanodevices (CRANN) & Advanced Materials and Bioengineering Research (AMBER) Centre, Trinity College Dublin, Dublin 2, Ireland.

[4] Materials Chemistry & Analysis Group, School of Chemistry and the Tyndall National Institute, University College Cork, Cork, Ireland.

[5] Department of Chemistry, Duke University, North Carolina, USA.

[†] Authors have contributed equally to this work.

**ABSTRACT:** Nanowire networks are promising memristive architectures for neuromorphic applications due to their connectivity and neurosynaptic-like behaviours. Here, we demonstrate a self-similar scaling of the conductance of networks and the junctions that comprise them. We show this behavior is an emergent property of any junction-dominated network. A particular class of junctions naturally leads to the emergence of conductance plateaus and a "winner-takes-all" conducting path that spans the entire network, and which we show corresponds to the lowest-energy connectivity path. These results point to the possibility of independently addressing memory or conductance states in complex systems and is expected to have important implications for neuromorphic devices based on reservoir computing.

The unique properties of nanoscale materials are well established. Currently, these properties are exploited through the integration of individual components (dots, wires, sheets) into devices or from the benefits derived from the assembly of these components into networks and composites. In each case the presence of surface layers – molecules, surfactants, polymers and native oxides – essential to stabilize these materials during synthesis and processing, represent barriers to physical integration and electrical connectivity. Thermal, mechanical and chemical processes have been employed to minimize these barriers and develop various applications based on metal nanowire networks (NWN) [1, 2] including flexible and transparent conductors [3, 4, 5, 6, 7, 8, 9, 10], energy storage [11, 12] and generator devices [13, 14, 15], sensors and memory devices [16, 17]. Nanoscale dielectric layers give rise to material-independent ubiquitous behaviors. For example, electrical stressing of junctions between oxide coated wires show resistive switching; polymer coated wires undergo controlled capacitive breakdown, whereas wires coated with semiconducting layers exhibit memristive-like properties [18, 19, 20, 21, 22, 23, 24]. Identical behaviors are found in planar metal-insulator-metal (MIM) structures [25, 26, 27], despite the different geometry and vastly different contact areas. These field-driven behaviors are pervasive on the nanoscale and have led to speculation about the formation of single "winner-takes-all" (WTA) conducting filaments (CF) that are believed to dominate conduction and memory performance [25, 28].

Here we demonstrate that it is possible to form WTA connectivity paths in macroscale nanowire networks. The existence of WTA paths is critical to establishing independently addressable memory or conductance states in complex systems. We describe the network properties necessary to establish a WTA path and the possibility of addressing nanoscale components within a macroscopic assembly without the need for direct contacts. To demonstrate this capacity, we first explore the relationship between the electrical behaviors of junctions formed between individual nanowires (NWs), with those of macroscopic assemblies of the same junctions present in random NWNs. We find that for both the increase in conductance scales identically with the current compliance limit used to assess the I-V characteristics of the system. This self-similar scaling holds for all nanomaterial systems studied and simulations reveal it is a property of any network where the junctions dominate transport (cf. Section 6 of the supplemental material for demonstration). Remarkably, we find for junctions with particular scaling properties, the associated macroscopic networks exhibit conductance plateaus at fractions of the quantum conductance level, $\Gamma_0 = 2e^2/h$. These conductance plateaus are indicative of the formation of a single WTA conducting path across the *entire* network. We demonstrate that WTA paths have the lowest energy of formation and are stable over a finite energy or input current range. Collectively, these results point to a capacity to self-select the lowest energy connectivity pathway within a complex random network, one that is robust and immune to perturbations. These observations are expected to have important implications for example in the area of neuromorphic devices based on reservoir computing [29, 30, 31] - neural network-based strategies for processing time-varying inputs that is highly effective for identification, prediction and classification tasks [32].

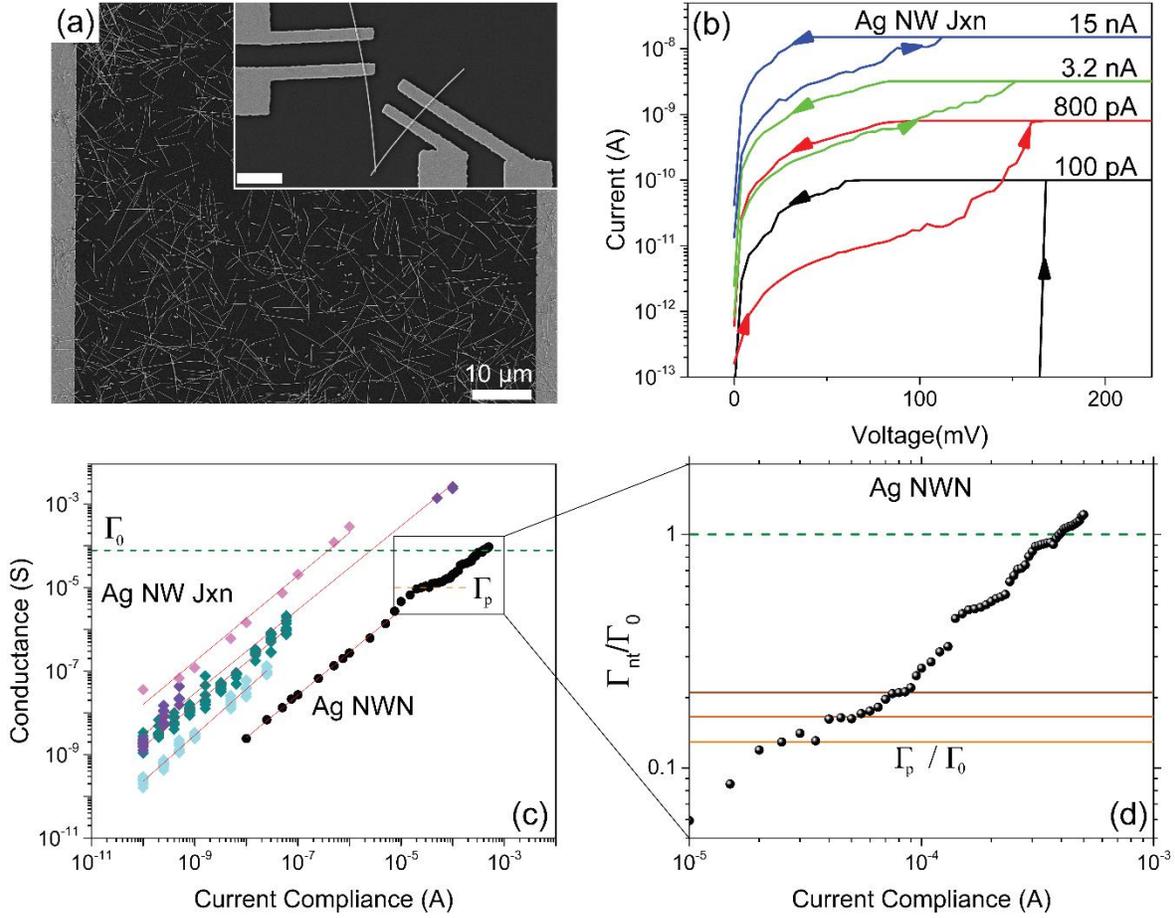

Figure 1: (a) Main panel shows a SEM image of an Ag NWN. All networks studied here have nearly the same wire density of ~0.3 NWs/µm$^2$. The inset depicts a SEM micrograph of a contacted Ag nanowire junction. (b) I-V curves for different programmed compliance currents for a single Ag nanowire junction. (c) Conductance plotted against the current compliance (log-log scale) for numerous Ag nanowire systems. Ag nanowire junctions (Ag NW Jxn) are represented by coloured diamonds. Black circles show the measurements for a 500 x 500 µm Ag NWN. The solid lines correspond to power law fit of these datasets. Both systems (Jxn and NWN) display a power-law dependence, however the Ag NWN shows a break from this trend near $\Gamma_0$ (the quantum of conductance) shown as the horizontal dashed green line. Measurement methods are detailed in Figures S1 and S2. (d) Zoom-in at the high current compliance range of the Ag NWN conductance ($\Gamma_{nt}$) taken from panel (c) and divided by $\Gamma_0$. This dataset was obtained using an extremely fine current compliance sampling in order to detect the numerous conductance plateaus marked by the solid horizontal lines. The green dashed line depicts the quantum of conductance. The orange horizontal line indicated the first conductance plateau ($\Gamma_p$) found at $\Gamma_p \approx 1 \times 10^{-5}$ S, which gives $\Gamma_p/\Gamma_0 \approx 1/8$. Subsequent plateaus (darker orange solid lines) appear at the $\approx 1/6$, and $\approx 1/5$.

## Results

Figure 1(a) shows an example of the single junction devices and nanowire networks used in this work. Four-probe contacts to single junction devices are established using e-beam lithography (EBL) and the contacts either side of the junction were electroformed prior to characterizing the junction itself [33]. Ag NWNs were spray deposited and contacted using a shadow mask that allowed for networks with sizes of 10 µm to 1000 µm to be tested. Figure 1(b) displays the data obtained from I-V sweeps on a single Ag nanowire junction at increasing current compliance levels ($I_c$). NWNs yielded similar curves but at larger applied voltages. Figure 1(c) compares the scaling behaviors of single nanoscale junctions with that of a network of junctions. Remarkably the conductance in both systems scale according to the same power-law (PL) $\Gamma = A\, I_c^{\alpha}$ where $\alpha$ is the scaling exponent and $A$ the PL prefactor. Table S1 in the supplemental material shows α and $A$ values obtained for a wide variety of systems. These results point to a similarity in the manner in which single junctions and networks of junctions become activated, i.e., a similarity between filament growth across a nanoscale junction and the formation of current pathways across the macroscopic network. The PL scaling behavior also holds true for junctions formed using a range of wire systems: Ni-NiO, core-shell Ag-TiO$_2$, and Cu-CuO, in addition to a range of planar electrochemical metallization memory (ECM) devices in the literature (cf. Figure S3 in the supplemental material). It is also important to note that all NWN samples studied in this work are 50 to 500 µm in size and experience current levels in the nA to µA range so as to avoid junction welding due to Joule heating.

    A second striking observation is the presence of plateaus in the network conductance below the quantum of conductance $\Gamma_0 = 2e^2/h$ [cf. Figure 1(d)]. Conductance plateaus were not observed in the case of single junctions as it is extremely difficult to control the conductance and

stabilize the formation of single channel CF within a single junction. Not all networks exhibit conductance plateaus, while some networks such as Ag NWNs always exhibit plateaus. We show below that conductance plateaus are observed only for networks with particular junction properties. For networks comprised of this special type of junction several plateaus are typically found, all at conductances below $\Gamma_0$ [cf. Figure 1(d)]. For each plateau, the network is stable over a range of input currents, and the connectivity within the network is not significantly affected by the additional power input. While quantum conductance effects has been previously observed in many systems, including a wide variety of conductive filament-based resistive random-access memory (CF RRAM) devices, percolating nanoparticle films, as well as single- and double-wall carbon nanotube networks [34, 35, 36, 37, 38], typically the observation of quantum conductance phenomena requires careful experimental design and precisely controlled electrical measurements. Evidently, networks facilitate detection even in macroscopic systems. We show below that the capacity of the network to absorb charge provides a ballast that stabilizes quantum conductance phenomena that occur in localized regions of a macroscopic network.

**Computational model.** To explain the origin of the self-similar scaling and the observation of quantized conductance in networks we performed simulations of network transport. To do so we introduce a "part-to-whole" scheme in which we make use of the experimental data gathered for individual junctions (part) to understand the collective behaviour of a network formed by those junctions (whole). The observed PL for the junction conductance ($\Gamma_j$) with current compliance ($I_c$) can be written as:

$$\Gamma_j = A_j I_c^{\alpha_j} \qquad (1)$$

where $A_j$ is a proportionality constant and $α_j$ is a positive exponent that fluctuates around 1 (cf. Table S1 in the supplemental material). By linking this power-law behaviour with the memristor charge-carrier drift model we show junctions are uniquely described by the set {$A_j$, $α_j$}, with $A_j$ related to the mobility of the diffusing species that build the filament in the dielectric and $α_j$ captures the non-linearity in the diffusion barrier (cf. Section 2 of the supplemental material). We assume the junction resistance itself is bounded but can have any value between a high resistance state (HRS) of $R_{off}$ ~ $10^4$ kΩ and a low resistance state (LRS) of $R_{on}$ = 12.9 kΩ (equivalent to $1/\Gamma_0$). These resistance cut-offs were chosen based on our measurements that show that some junctions can in fact reach $\Gamma_0$ at sufficiently high currents. A graphical representation of our Power-Law plus Cut-offs (PL+C) junction-model can be found in Figure S6 of the supplemental material.

Junctions modelled within PL+C fall into three different types: sub-linear ($α_j < 1$), linear ($α_j = 1$), or supra-linear ($α_j > 1$) each associated with a different strengthening rate ($v_j$) at which its conductance increases with the current:

$$v_j = \frac{d\Gamma_j}{dI_c} = A_j \alpha_j I_c^{\alpha_j - 1} \qquad (2).$$

The exponent $α_j$ determines the rate with which the conductance of an individual junction changes; whether it speeds-up ($α_j > 1$), stays constant ($α_j = 1$), or slows down ($α_j < 1$) in response to increasing current. The prefactor $A_j$ determines how quickly a junction reaches the LRS. Once the junction reaches its LRS, it becomes a regular resistor with fixed resistance of $R_{on}$ (cf. Figure S5 in the supplemental material).

We now demonstrate that the transport properties of NWNs comprised of junctions modelled within PL+C naturally give rise to emergent behaviours such as self-similarity, conductance plateaus and the selective formation of conducting paths that are immune to input perturbations. To describe such NWNs we exploit our multi-nodal representation [39, 40] which accounts for all resistances, including junction resistances ($R_j$) and inner wire resistances ($R_{in}$). The latter are fixed quantities determined as $R_{in} = \rho l/A_c$ where ρ is the wire resistivity, $l$ its segment length, and $A_c$ its cross-sectional area. An initial amount of current I = $I_0$ is sourced at the electrodes and Kirchhoff's circuit equations written in matrix form, $\widehat{M}_R \widehat{U} = \widehat{I}$, are solved where $\widehat{M}_R$ is the resistance matrix (its elements are the inverse of the resistance), and the vectors $\widehat{U}$ and $\widehat{I}$ are the potential at each circuit node and the current injected/drained out of the device, respectively. The current flowing through each junction is mapped onto a pair of voltage nodes (n,m) and calculated using

$$I_{n,m} = \frac{|U_n - U_m|}{R_j^{n,m}} \qquad (3).$$

For the first iteration, $R_j^{n,m}$ = $R_{off}$ ∀ (n,m) internode pairs. Once $I_{n,m}$ is determined for all junctions, their new conductance state is obtained using the same functional as in equation (1), i.e.

$$\Gamma_j^{n,m} = A_j (I_{n,m})^{\alpha_j} \qquad (4).$$

After updating the conductance of all junctions, the total current sourced on the electrodes is incremented as $I \rightarrow I + \Delta I$ and the whole procedure of calculating $\Gamma_j^{n,m}$ takes place recursively until I reaches a predefined maximum value of $I_{max}$. Note that all junctions are subjected to the same cut-off limits set by the [$R_{off}$, $R_{on}$] window. The sheet conductance of the network ($\Gamma_{nt}$) is then calculated recursively for each current value I and this outcome is used to compare with the experimental curves of $\Gamma_{exp}$ vs $I_c$ [cf. Figure 1(c)]. The workflow of the algorithm can be found in the supplemental material (cf. Figure S6).

Figure 2 shows conductance versus current simulation for an actual Ag NWN of density 0.49 NWs/μm². The micrograph image of this network is shown in the supplemental material (cf. Figure S7). Each curve represents a distinct set of {$A_j$, $α_j$} values and in each case four distinct conducting regimes are observed: (i) OFF-threshold (OFF), (ii) transient growth (TG), (iii) power-law (PL), and (iv) post-power-law (PPL). In the OFF-threshold regime, all junctions are in the OFF-state and the network is not distributing enough current to improve their resistances. At a certain critical current, the conductance of the network increases in a nonlinear fashion as junctions begin to improve their resistances (TG regime). Note that the OFF→TG crossover current depends on the choice of {$A_j$, $α_j$} and consequently on the strengthening rate $v_j$. During the PL stage, we fit $Γ_{nt} = A_{nt} I^{α_{nt}}$ onto the numerical curves to obtain the network phase space {$A_{nt}$, $α_{nt}$}, which we then compare with the junction phase space {$A_j$, $α_j$}. The results are presented in Table 1 and from which we find that $α_{nt} = α_j$ in each case. Small discrepancies are observed for larger values of $A_j$ but overall, our simulations demonstrate a one-to-one correspondence between $α_j \leftrightarrow α_{nt}$, consistent with self-similar behaviour found in the experiments.

Table 1: {$A_{nt}$, $α_{nt}$} values for networks obtained by fitting the power law $Γ_{nt} = A_{nt} I^{α_{nt}}$ onto the curves of Figure 2. All junctions in a given network are set to have the same prefactor and exponent except for the heterogeneous case in which a narrow dispersion was induced in the exponents using a truncated normal distribution with mean value of $\langle α_j \rangle$. Note the strong correlation between $α_{nt}$ and $α_j$.

| $A_j$ | 0.01 | 0.05 | 0.1 | 0.5 |
|---|---|---|---|---|
| $α_j = 0.9$ | {0.0027, 0.892} | {0.0133, 0.896} | {0.0266, 0.9} | {0.1407, 0.925} |
| $α_j = 1.0$ | {0.0025, 1.0} | {0.0125, 1.0} | {0.0251, 1.0} | {0.13071, 1.024} |
| $α_j = 1.1$ | {0.0024, 1.115} | {0.0125, 1.115} | {0.0251, 1.113} | {0.13941, 1.159} |
| $\langle α_j \rangle = 1.05$ | {0.0025, 1.054} | {0.0125, 1.049} | {0.0251, 1.051} | {0.1323, 1.071} |

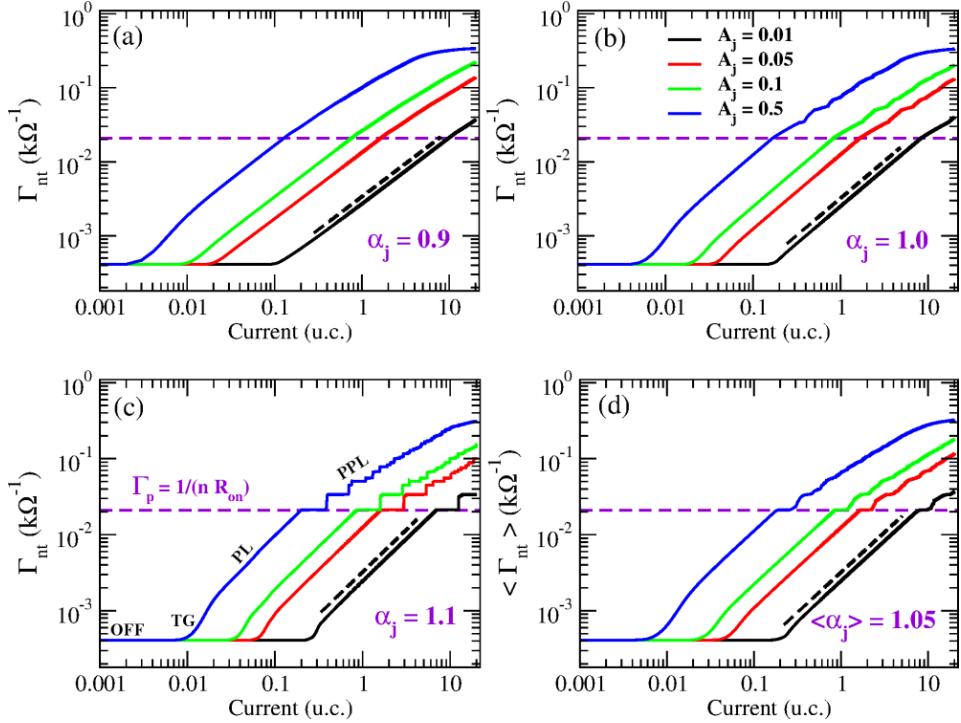

Figure 2: Conductance versus current plots taken for an image processed Ag NWN of wire density of 0.49 NWs/μm$^2$ (cf. Figure S7 in the supplemental material). Currents are expressed in units of current (u.c.). The results were taken for distinct values of A$_j$ and for exponents of (a) $\alpha_j = 0.9$, (b) $\alpha_j = 1$, (c) $\alpha_j = 1.1$, and (d) $\langle \alpha_j \rangle = 1.05$. In the latter, a narrow dispersion was induced in the exponents using a normal distribution with $\langle \alpha_j \rangle = 1.05$, σ = 0.1 and truncated at [1.0,1.1]. Black dashed lines illustrate the power-law fittings that determined $\{A_{nt}, \alpha_{nt}\}$. Results for all fittings are presented on Table 1. Horizontal dashed lines mark the conductance of the first path formed in the network containing n junctions at their optimal state R$_{on}$. This conductance level is given by $\Gamma_p = 1/(nR_{on})$ and for this particular network, n=4. A distinction between the four distinct transport regimes discussed on the main text is depicted on panel (c): (OFF) OFF-threshold, (TG) transient growth, (PL) power law, and (PPL) post-power-law.

Figure 2 also points to a rich behaviour in the PPL regime. Whereas single junctions simply transform into ordinary resistors, networks are highly interconnected and continue to evolve in response to the applied current. The actual behaviour depends on the value of α$_j$. For sub-linear junctions as shown in Figure 2(a) ($\alpha_j < 1$), the curves vary smoothly with an almost imperceptible change in slope at higher current marking the transition PL → PPL. When α$_j$ = 1 [cf. Figure 2(b)], the curves show fine structures and finally when $\alpha_j > 1$ discrete conductance plateaus emerge that are strikingly similar to those found experimentally [cf. Figure 1(c)]. Figure 2(d) considers a small dispersion in $\alpha_j$ and an average conductance curve was obtained for a configurational ensemble containing 10 sets of $\alpha_j$ distributions used to describe the junctions on the network skeleton (see figure caption of Figure 2). One can see that apart from the overall step-like conductance behaviour being smoothen out, a robust plateau and other numerous discontinuities are preserved in all curves showing that step-like features in the conductance curves are fingerprints of NWNs composed of supra-linear junctions. We note that Ag junctions considered here are supra-linear with $\alpha_j \sim 1.05 - 1.1$.

To explore the precise origin of plateaus in networks with supra-linear junctions, we calculate the amount of current flowing through each wire segment ($I_s$) and plot these data in contour maps in Figure 3. The network segment-skeleton that serves as a template for these current maps is shown in Figure S7 in the supplemental material. Current-contour maps were obtained at distinct source-current values to capture the evolution of the conductance state of the network in the TG, PL, and PPL regimes. These states are identified in the top-left curve by different shaped symbols (square, star, triangle and circle). The corresponding current maps are displayed in the coloured panels in Figure 3 and provide a graphical representation of the distinct conducting pathways that emerge naturally as the source current is increased. In the TG regime, the limited input-current spreads across the network to probe the most efficient way to transmit the sourced current. As the input-current increases, the network could, in principle, keep availing of the hundreds of path combinations evidenced in the TG current map to transmit the sourced signal but instead, it selects the "easiest-conducting-path", the WTA path. Once the WTA path is fully optimized (with all its junctions at the LRS when I = 2.25 u.c.), the network becomes temporarily

Ohmic, reflected by the appearance of the first conductance plateau in the $\Gamma_{nt} \times I$ curve. Since the inner resistance of the metal wires is negligible, the conductance of this first plateau $(\Gamma_p)$ is found at approximately $\Gamma_p = 1/(nR_{on}) = \Gamma_0/n$, where n is the number of junctions along the path. This means that each junction along the path behaves as a single conduction channel with an effective resistance of $R_{on} = 1/\Gamma_0$. As more current is pushed through the network, other paths are opened in a discrete fashion leading to discontinuous jumps and additional conductance plateaus. Three well-defined paths (two partially superimposed) can be seen in the contour plot of Figure 3 with sourced current of I = 7.0 u.c. A similar current map for a different network system is provided in the supplemental material (cf. Figure S8) to confirm the generality of this behaviour. These findings are consistent with the experimentally measured conductance-plateaus being located at fractions of $\Gamma_0$ [cf. Figure 1(d)]. Note that these conductance-plateaus can be observed only when extremely fine current steps are used in the measurements.

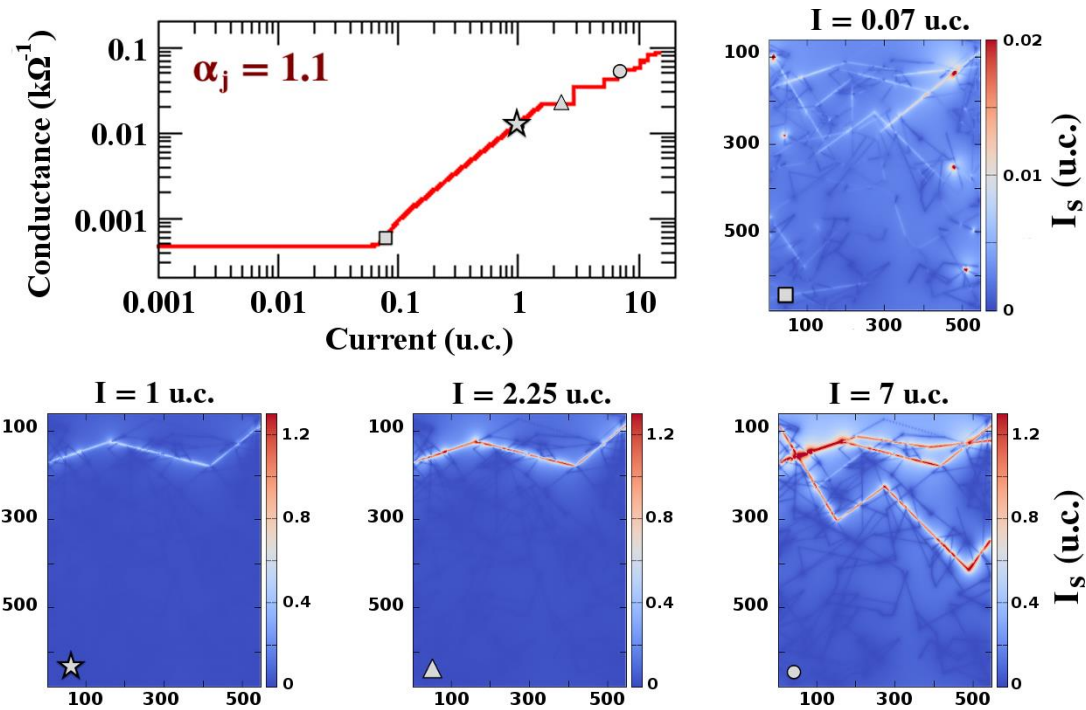

Figure 3: (Top-left panel) The same conductance versus current curve shown in Figure 2 for the Ag NWN template depicted in Figure S5 of the supplemental material. The junction characteristics are set at $A_j =$

0.05 and $\alpha_j = 1.1$. The symbols mark points in the curves in which current colour maps were taken. (Contour panels) Current colour maps calculated over each wire segment ($I_s$) of the Ag NWN. Snapshots were taken for four current values specified on the top of each current map and distinguished by the symbols: square (TG), star (PL), triangle and circle (both set in the PPL regime). In particular, the PPL state at I = 2.25 u.c. is located at the first conductance plateau, $\Gamma_p = \Gamma_0/4$. Animations revealing the complete evolution of the network in response to the current source, junction optimization of the top-3 paths of least-resistance, and current-segment maps are provided in the supplemental material (cf. Section 7 and Figure S13 for animation description).

The actual WTA path chosen by the NWN is controlled by several factors. The random nature of the network skeleton serves as a first selection mechanism for the propagation of current as it eliminates spatial redundancies. The second selection mechanism relates to the memristive character of the junctions, in particular the value of $\alpha_j$ and consequently the strengthening rate $v_j$. Figure S9 in the supplemental material depicts the current maps for all four conducting regimes addressed in Figure 3 but it also includes the results for the same network interconnected by linear and sub-linear junctions. When $\alpha_j \leq 1$, the network starts its dynamics by selecting two superposing paths that initially have nearly equal least-resistance levels. These two paths are strengthened throughout the PL regime until other paths start to emerge gradually in the PPL stage. The situation changes dramatically when $\alpha_j > 1$ with the network selecting a single conducting path in the PL regime. This is a direct consequence of some junctions being optimized faster than others. Under these conditions, the strengthening rate of the junctions described by equation (2) increases with the current. This means that junctions favoured by the network topology will improve at a faster pace whereas those that are hindered by the wire assembly will be delayed in the race for optimal conductance state. This mechanism naturally leads to a WTA outcome similar to that employed in supervised competitive learning in recurrent neural networks [41, 42].

The introduction of dispersion in the exponents further drives the network in choosing a single path in comparison to the idealized case in which all junctions have the same characteristics (cf. Figure S10 in the supplemental material). Even paths that nominally have the same initial conductance rapidly differentiate when dispersion is included. Mixing junctions of distinct strengthening rates is a third selection mechanism for the network with some junctions evolving faster than others. The fact that these junctions with slightly different properties are embedded in a highly disordered NW template makes each one of them a unique building block of the network and so redundancies in the propagation of current and path formation are readily eliminated. Deviations from the PL behaviour can occur however when the inner resistance of the wires starts to play an important role. To demonstrate this effect, we artificially increased the resistivity of the wires while keeping the junction resistances fluctuating in the same [$R_{off}$, $R_{on}$] range. These results are shown in Figure S12 in the supplemental material and they indicate that a clean PL scaling behaviour and a conducting WTA state can only be found for systems in which $R_{in} \ll R_j$. Other sample characteristics such as NW geometrical aspects, network density, width of the devices, contact resistance (cf. Figure S11 in the supplemental material), and electrode separation are also found to affect the details of PL dynamics and the formation of WTA paths in random NWNs because they all have an impact on the network connectivity [43, 44, 45]. That said, any network for which $R_{in} \ll R_j$ and $\alpha_j > 1$, will exhibit a WTA path.

We now demonstrate that the self-selective behaviour of the network is energy-efficient. The power consumed by the network is given by $P_{nt} = I^2/\Gamma_{nt}$ and in the PL regime since $\Gamma_{nt} = A_{nt} I^{\alpha_{nt}}$, we can write

$$P_{nt} = \frac{I^{2-\alpha_{nt}}}{A_{nt}} \qquad (7)$$

so that the power varies non-quadratically with the current. Since the network selects a path with the largest exponent $\alpha_{nt}$, equation (7) insures that the establishment of this path dissipates the lowest possible power. As more current is pushed through, the network enters into the PPL regime and a conductance plateau is formed. Since $\alpha_{nt} = 0$ in the plateau region, $P_{nt} = I^2/A_{nt}$ and hence the network becomes Ohmic. In the limit of sufficiently high currents, and after the emergence of successive conductance plateaus, one expects that almost 100% of the junctions will be fully optimized and the network reaches saturation with $\Gamma_{nt} \rightarrow$ constant and $P_{nt} \propto I^2$.

To confirm the simulation outcomes and to directly visualise WTA paths, passive voltage contrast (PVC) SEM images were acquired in Ag NWNs by grounding the electrodes at selected points during a conductance versus current-compliance measurement. Wires that are connected to a grounded electrode through electrical activation appear darker, whereas disconnected or less well-connected wires appear brighter, so that a PVC image can provide a qualitative comparison of the electrical connectivity within a given network. Figure 4(a) shows the unperturbed network with no contrasting wires in the image. The current compliance was then gradually ramped up and at $I_c = 50 \, nA$ in Figure 4(b) we observe the emergence of WTA path that is about to bridge the bottom electrode. On increasing the current, the path is established and then reinforced and at $I_c = 500 \, nA$ shown in Figure 4(c) those wires taking part in the conduction dominate the contrast. As the compliance current is increased further the path is strengthened as evidence by the increased relative contrast, until finally secondary paths begin to emerge in Figure 4(d) as two extra percolating paths emerge from the top electrode. PVC measurements performed on larger NWN samples are shown in the supplemental material (cf. Figure S14). They help visualise the behaviour in the TG and PPL regimes prior and after the WTA has been chosen. In the TG regime (recorded following a current of few pA) large areas of the network show contrast consistent with the simulations in Figure 3 during which the entire network is uniformly probed prior to selecting the WTA path. Figure S14(b) shows the presence of multiple conducting paths in the PPL regime.

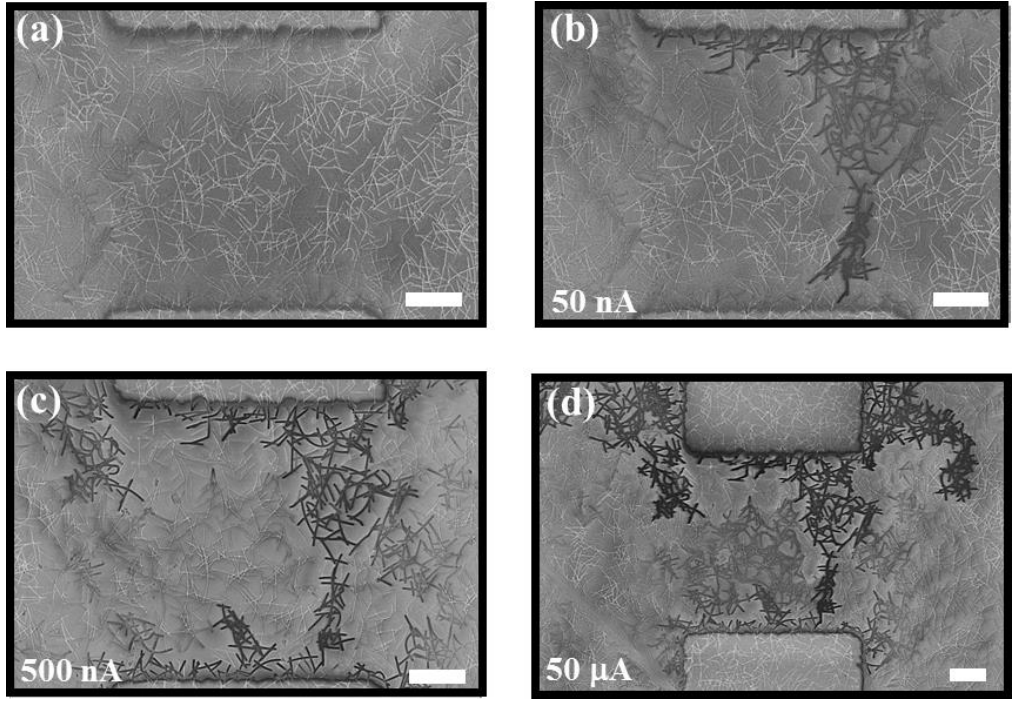

Figure 4: (a) SEM image of an unperturbed Ag NWN of dimensions 100 x 100 μm. (b-d) PVC images of the same network taken during I-V sweeps with limiting current compliances of (b) 50 nA, (c) 500 nA, and (d) 50 μA. Note it is not possible to directly compare the contrast observed in different networks or even that observed in the same network imaged under different conditions. Current levels are written on the respective panels. White scale bars correspond to 2 μm. The respective conductance values measured from the I-V curves are (b) $\Gamma_{nt} = 1.1 \times 10^{-7}$ S, (c) $\Gamma_{nt} = 1.0 \times 10^{-6}$ S, and (d) $\Gamma_{nt} = 9.6 \times 10^{-6}$ S.

In summary, we have described the junction and network properties necessary to establish WTA paths in random nanowire networks. PL behaviour is expected for any network in which the conductance is controlled by the junction properties so that self-similar scaling of junctions and networks is a natural property of junction dominated networks. However, networks comprised of junctions that strengthen under current flow ($\alpha_j > 1$) lead to the development of WTA paths that exhibit characteristic conductance plateaus that are stable over a current compliance range. WTA paths represent the lowest possible energy paths and enable the establishment of independently addressable memory or conductance states within complex network systems. Further work is needed to understand how to engineer the properties of junctions to better satisfy the $\alpha_j > 1$ requirement. These findings will help in the development of novel hardware neural network systems with brain-inspired architectures for cognitive signal-processing, decision-making systems and ultimately neuromorphic computing applications.

## Methods

**Measurements and materials:** Further information on the nanowires used in this study, including TEM/SEM images, details of I-V and I-t experiments (t being the time) can be found in Figures S1 and S2 in the supplemental material. Comparisons between scaling of CBRAM (Conductive Bridging RAM) literature data and experimental data from various NW systems can also be found in the supplemental material (cf. Figure S3). Experiments were carried out on P-type Silicon wafers (University Wafer) with a 300 nm thermally grown $SiO_2$; nanowire solutions were dropcast onto substrates pre-patterned by UV lithography. Single, crossed wires, and isolated NWNs were fabricated using previously reported techniques [39]. Larger NWNs were spray-deposited and contacted using a shadow mask.

**Electron microscopy and electrical measurements:** Scanning electron microscopy (SEM) images and EBL was performed on a Zeiss Supra FEG-SEM. Transmission electron microscopy (TEM) images were acquired using a FEI TITAN TEM. Electrical measurements were carried out at ambient conditions using two set ups. A Keithley 4200 Semiconductor characterization station was primarily used for electrical measurements. Experiments were also undertaken on a Keithley 3450 and 6450 femtoamp paired with custom LabView interface.

**Passive Voltage Contrast (PVC) experiments:** A Zeiss Ultra FEG-SEM paired with a Keithley 2400 and Kleindiek Nanotechnik probing system was used to performed PVC imaging [46]. The secondary electrons, produced by a low energy electron beam (2-4 kV) were used to visualize parts of the network electrically activated. Small apertures were also used to improve the quality of the image reducing the background contrast.

Corresponding Author:
*E-mail: jboland@tcd.ie.

**Acknowledgements:**
The authors wish to acknowledge funding from the European Research Council under Advanced Grant 321160. This publication has emanated from research supported in part by a research grant from Science Foundation Ireland (SFI) AMBER Centre under Grant Number SFI/12/RC/2278. The facilities and staff at the Advanced Microscope Laboratory at Trinity College Dublin are greatly acknowledged for their support, as is the Research IT Unit (formerly TCHPC) at Trinity College Dublin and The Irish Centre for High-End Computing (ICHEC) for computational resources. The authors which to thank Anurag Gupta for programming the LabView Software used for data acquisition, and developing custom test modules for the Keithley SCS-4200.

**Author Contributions:**
H.G.M. co-wrote the paper, and along with A.T.B. performed experiments on individual NWs, junctions. F.N. performed measurements on NWN samples and passive voltage experiments. C.G.R. co-wrote the paper, and along with C.O'C. developed the computational model and ran the simulations. S.B. fabricated Ag-TiO$_2$ NWs, and P.F. fabricated Cu NWs. M.S.F. developed the computational model; and J.J.B. led overall effort and co-wrote the paper. All authors discussed and commented on the manuscript and on the results.

For additional information see the supporting material available.

**REFERENCES:**

[1] Lagrange M, Langley D, Giusti G, Jiménez C, Bréchet Y, Bellet D. Optimization of silver nanowire-based transparent electrodes: effects of density, size and thermal annealing. *Nanoscale* **7**, 17410-17423 (2015).

[2] Large MJ, Cann M, Ogilvie SP, King AAK, Jurewicz I, Dalton AB. Finite-size scaling in silver nanowire films: design considerations for practical devices. *Nanoscale* **8**, 13701 (2016).

[3] Lee J, et al. Room-temperature nanosoldering of a very long metal nanowire network by conducting-polymer-assisted joining for a flexible touch-panel application. *Advanced Functional Materials* **23**, 4171-4176 (2013).

[4] Suh YD, et al. Random nanocrack, assisted metal nanowire-bundled network fabrication for a highly flexible and transparent conductor. *RSC Advances* **6**, 57434-57440 (2016).

[5] Moon H, Won P, Lee J, Ko SH. Low-haze, annealing-free, very long Ag nanowire synthesis and its application in a flexible transparent touch panel. *Nanotechnology* **27**, 295201 (2016).

[6] Hong S, et al. Highly stretchable and transparent metal nanowire heater for wearable electronics applications. *Advanced Materials* **27**, 4744-4751 (2015).

[7] Lee P, et al. Highly stretchable and highly conductive metal electrode by very long metal nanowire percolation network. *Advanced Materials* **24**, 3326-3332 (2012).

[8] Han S, et al. Nanorecycling: monolithic integration of copper and copper oxide nanowire network electrode through selective reversible photothermochemical reduction. *Advanced Materials* **27**, 6397-6403 (2015).

[9] Han S, et al. Fast plasmonic laser nanowelding for a Cu-nanowire percolation network for flexible transparent conductors and stretchable electronics. *Advanced Materials* **26**, 5808-5814 (2014).

[10] Rathmell AR, Nguyen M, Chi M, Wiley BJ. Synthesis of oxidation-resistant cupronickel nanowires for transparent conducting nanowire networks. *Nano Letters* **12**, 3193-3199 (2012).

[11] Lee H, et al. Highly stretchable and transparent supercapacitor by Ag–Au core–shell nanowire network with high electrochemical stability. *ACS Applied Materials & Interfaces* **8**, 15449-15458 (2016).

[12] Moon H, Lee H, Kwon J, Suh YD, Kim DK, Ha I, Yeo J, Hong S, Ko SH. Ag/au/polypyrrole core-shell nanowire network for transparent, stretchable and flexible supercapacitor in wearable energy devices. *Scientific Reports* **7**, 41981 (2017).

[13] Jeong CK, et al. A Hyper-Stretchable Elastic-Composite Energy Harvester. *Advanced Materials* **27**, 2866-2875 (2015).

[14] Chang I, Park T, Lee J, Lee MH, Ko SH, Cha SW. Bendable polymer electrolyte fuel cell using highly flexible Ag nanowire percolation network current collectors. *Journal of Materials Chemistry A* **1**, 8541-8546 (2013).

[15] Chang I, et al. Performance enhancement in bendable fuel cell using highly conductive Ag nanowires. *International Journal of Hydrogen Energy* **39**, 7422-7427 (2014).

[16] Kim KK, et al. Highly sensitive and stretchable multidimensional strain sensor with prestrained anisotropic metal nanowire percolation networks. *Nano Letters* **15**, 5240-5247 (2015).

[17] Du H, et al. Engineering silver nanowire networks: from transparent electrodes to resistive switching devices. *ACS Applied Materials & Interfaces* **9**, 20762-20770 (2017).

[18] Manning HG, Biswas S, Holmes JD, Boland JJ. Nonpolar resistive switching in Ag@TiO$_2$ core-shell nanowires. *ACS Applied Materials & Interfaces*, DOI: 10.1021/acsami.7b10666 (2017).

[19] Nirmalraj PN, et al. Manipulating connectivity and electrical conductivity in metallic nanowire networks. *Nano Letters* **12**, 5966-5971 (2012).

[20] Ielmini D, Cagli C, Nardi F, Zhang Y. Nanowire-based resistive switching memories: devices, operation and scaling. *Journal of Physics D: Applied Physics* **46**, 074006 (2013).

[21] O'Kelly CJ, Fairfield JA, Boland JJ. A single nanoscale junction with programmable multilevel memory. *ACS Nano* **8**, 11724-11729 (2014).

[22] O'Kelly CJ, Fairfield JA, McCloskey D, Manning HG, Donegan JF, Boland JJ. Associative enhancement of time correlated response to heterogeneous stimuli in a neuromorphic nanowire device. *Advanced Electronic Materials* **2**, 1500458 (2016).

[23] Fairfield JA, Ritter C, Bellew AT, McCarthy EK, Ferreira MS, Boland JJ. Effective electrode length enhances electrical activation of nanowire networks: experiment and simulation. *ACS Nano* **8**, 9542-9549 (2014).

[24] Fan Z, Fan X, Li A, Dong L. Resistive switching in copper oxide nanowire-based memristor. 2012 *12th IEEE Conference on Nanotechnology* (IEEE-NANO), DOI: 10.1109/NANO.2012.6322196.

[25] Jeong DS, et al. Emerging memories: resistive switching mechanisms and current status. *Reports on Progress in Physics* **75**, 076502 (2012).

[26] Ielmini D, Waser R. *Resistive Switching: From Fundamentals of Nanoionic Redox Processes to Memristive Device Applications*. John Wiley & Sons (2015).

[27] Zahari F, Hansen M, Mussenbrock T, Ziegler M, Kohlstedt, H. Pattern recognition with TiO$_x$-based memristive devices. *AIMS Materials Science* **2**, 203-216 (2015).

[28] Celano U, Giammaria G, Goux L, Belmonte A, Jurczak M, Vandervorst W. Nanoscopic structural rearrangements of the Cu-filament in conductive-bridge memories. *Nanoscale* **8**, 13915-13923 (2016).

[29] Demis EC, et al. Nanoarchitectonic Atomic switch networks for unconventional computing. *Japanese Journal of Applied Physics* **55**, 1102B2 (2016).

[30] Sillin HO, et al. A theoretical and experimental study of neuromorphic atomic switch networks for reservoir computing. *Nanotechnology* **24**, 384004 (2013).

[31] Avizienis AV, et al. Neuromorphic Atomic Switch Networks. *PLoS ONE* **7**, e42772 (2012).

[32]  While the connectivity structure of the network or reservoir remains fixed, the nodes (the junctions in the case of NWNs) evolve dynamically in response to input signals and collectively define the internal state of the reservoir. This serves to map lower-dimensional input signals onto outputs of higher dimensions, which are then examined by an external readout function.

[33]  Bellew AT, Manning HG, Rocha CG, Ferreira MS, Boland JJ. Resistance of single Ag nanowire junctions and their role in the conductivity of nanowire networks. *ACS Nano* **9**, 11422-11429 (2015).

[34]  Tappertzhofen S, Valov I, Waser R. Quantum conductance and switching kinetics of AgI-based microcrossbar cells. *Nanotechnology* **23**, 145703 (2012).

[35]  Jameson JR, et al. Quantized conductance in conductive-bridge memory cells. *IEEE Electron Device Letters* **33**, 257-259 (2012).

[36]  Ohno T, Hasegawa T, Tsuruoka T, Terabe K, Gimzewski JK, Aono M. Short-term plasticity and long-term potentiation mimicked in single inorganic synapses. *Nature Materials* **10**, 591-595 (2011).

[37]  Baxendale M, Melli M, Alemipour Z, Pollini I, Dennis T. Quantum conductance in single- and double-wall carbon nanotube networks. *Journal of Applied Physics* **102**, 103721 (2007).

[38]  Sannicolo T, et al. Direct imaging of the onset of electrical conduction in silver nanowire networks by infrared thermography: evidence of geometrical quantized percolation. *Nano Letters* **16**, 7046 (2016).

[39]  Rocha CG, et al. Ultimate conductivity performance in metallic nanowire networks. *Nanoscale* **7**, 13011-13016 (2015).

[40]  Fairfield JA, Rocha CG, O'Callaghan C, Ferreira MS, Boland JJ. Co-percolation to tune conductive behaviour in dynamical metallic nanowire networks. *Nanoscale* **8**, 18516-18523 (2016).

[41]  Maass W. On the computational power of winner-take-all. *Neural Computation* **12**, 2519-2535 (2000).

[42]  Fang Y, Cohen MA, Kincaid TG. Dynamics of a winner-take-all neural network. *Neural Networks* **9**, 1141-1154 (1996).

[43]  O'Callaghan C, Rocha CG, Manning HG, Boland JJ, Ferreira MS. Effective medium theory for the conductivity of disordered metallic nanowire networks. *Physical Chemistry Chemical Physics* **18**, 27564-27571 (2016).

[44]  Kumar A, Vidhyadhiraja NS, Kulkarni GU. Current distribution in conducting nanowire networks. *Journal of Applied Physics* **122**, 045101 (2017).

[45]  Žeželj M, Stanković I. From percolating to dense random stick networks: conductivity model investigation. *Physical Review B* **86**, 134202 (2012).

[46]  Gemmill Z, Durbha L, Jacobson S, Gao G, Weaver K. SEM and FIB passive voltage contrast. *Microelectronic Failure Analysis Desk Reference*, 431-437 (2004).

# Supplemental Material: "Emergence of Winner-takes-all Connectivity Paths in Random Nanowire Networks"

Hugh G. Manning[1,3,†], Fabio Niosi[1,3,†] Claudia Gomes da Rocha[2,3,†], Allen T. Bellew[1,3], Colin O'Callaghan[2,3], Subhajit Biswas[4], Patrick Flowers[5], Ben J. Wiley[5], Justin D. Holmes[4], Mauro S. Ferreira[2,3], and John J. Boland[1,3*]

[1] School of Chemistry, Trinity College Dublin, Dublin 2, Ireland.

[2] School of Physics, Trinity College Dublin, Dublin 2, Ireland.

[3] Centre for Research on Adaptive Nanostructures and Nanodevices (CRANN) & Advanced Materials and Bioengineering Research (AMBER) Centre, Trinity College Dublin, Dublin 2, Ireland.

[4] Materials Chemistry & Analysis Group, School of Chemistry and the Tyndall National Institute, University College Cork, Cork, Ireland.

[5] Department of Chemistry, Duke University, North Carolina, USA.

[†] Authors have contributed equally to this work.

**1. Additional experimental data and information about the material samples**

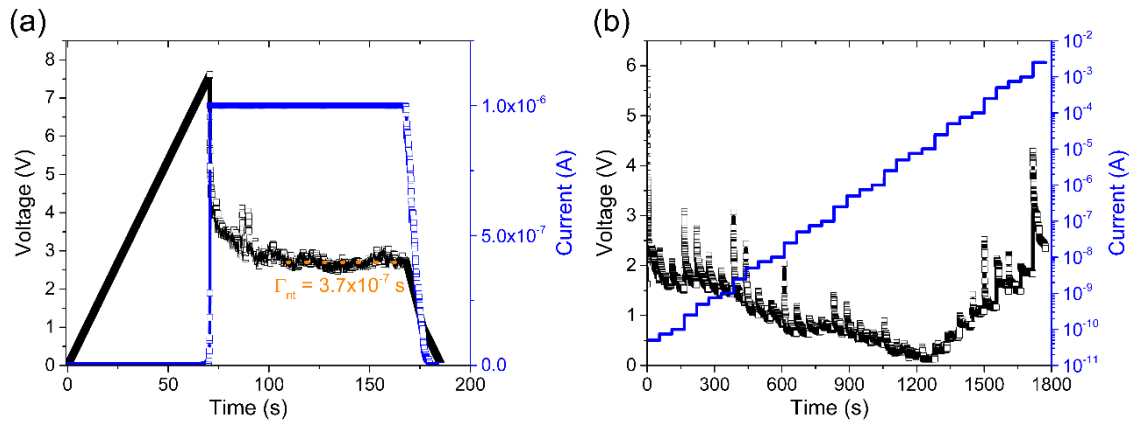

Figure S1: Measured voltage and current plots for a 200 x 200 µm Ag NWN. (a) An increasing voltage is sourced across the Ag NWN as shown by the increasing black line, as a conducting pathway is formed through the network and current flows the device quickly reaches a preset current compliance (1 µA, shown in blue) at 7.7 V. The subsequent drop in the measured voltage is required to maintain the 1 µA compliance level. The conductance of the network, $\Gamma_{nt}$, is taken as an average over the region before the current returns from the compliance limit, shown by the orange dashed line. This measurement was repeated at increasing current compliance levels resulting in the conductance vs current compliance plots shown in Figure 1(c) of the main text. (b) Conductance plateaus found in Figure S4(a-b) were obtained using a current step measurement, where a constant current is sourced in a stepwise manner (shown in blue) with the voltage required to maintain this current captured by the black data points. $\Gamma_{nt}$ was obtained for each compliance value as an average after the initial spike and decay of the applied voltage. The minimum applied voltage at ~1200 s represents the WTA path and plateau region where the network dissipates the lowest power possible. The subsequent increase in voltage is required to form additional pathways.

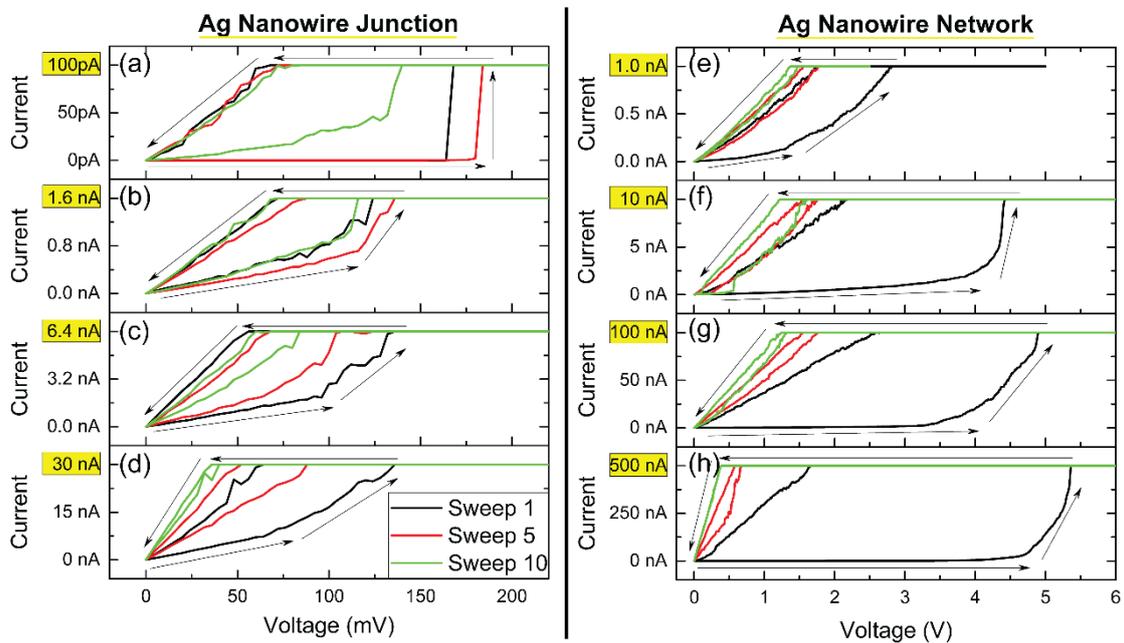

Figure S2: Experimentally obtained I-V curves for 10 successive sweeps (sweep 1, 5 and 10 shown in black, red and green, respectively) for a single Ag NW junction (a-d) and a 500 µm Ag NWN (e-h) at increasing current compliance levels. Large hysteresis loops are observed for low current compliance levels (a-b) with hysteresis loops collapsing after an increasing number of sweeps when larger currents are passed through the device (c-d). The collapse of the hysteretic loops is more pronounced in the Ag NWNs (e-h). After the initial sweep (black curve) a much higher leakage current is collected for the subsequent measurements. In all cases, between each set of 10 sweeps, the system was allowed to relax for a short period of time (~ 1 minute) which causes the hysteresis loops to reopen.

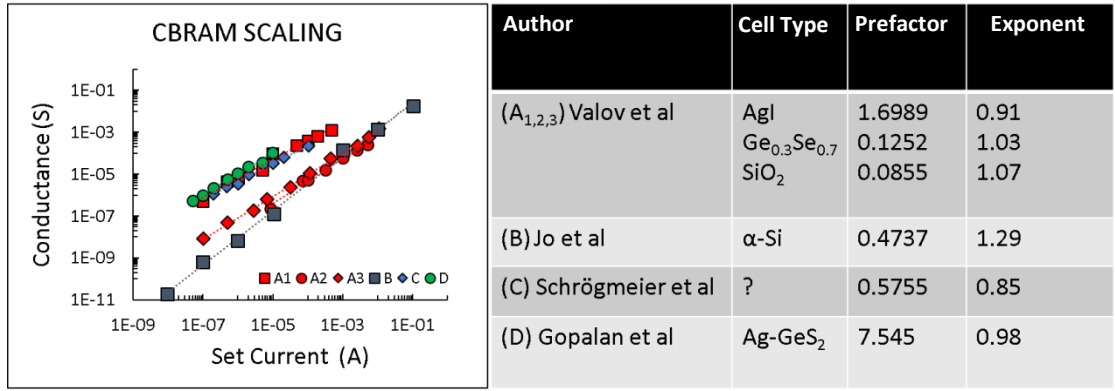

Figure S3: (Left panel) CBRAM scaling curves of thin film cells found in the literature. Differing materials show similar exponents all around one. (Right panel) Table of material systems containing prefactors and exponents taken from Valov et al. [1], Jo et al. [2], Schrögmeier et al. [3], and Gopalan et al. [4] as depicted on the left panel.

**Summary of Nanowires Systems Studied in this Work**

Ag Nanowires

Single crystalline pentagonal twinned Ag nanowires (Seashell Technology) with a thin polymeric (PVP) coating from synthesis were used in all experiments. TEM confirms the crystal structure of the nanowire and the PVP coating (1-2 nm thick).

Ag-TiO$_2$ Nanowires

Ag-TiO$_2$ NWs were synthesized using a solvothermal growth method as reported by Du et al. [5]. These wires have an Ag core (136 nm in diameter) with an amorphous TiO$_2$ coating (40 nm). EDX and TEM confirm the wire structure and the amorphous crystallinity of the TiO$_2$ shell.

Cu Nanowires

Cu nanowires were synthesized using the method described by Rathmell et al. [6] Cu nanowires develop a thin oxide layer through nanowire processing and exposure to ambient conditions. TEM analysis of the nanowires shows the single crystalline core with oxide coating (5.3 nm).

Ni Nanowires

Ni Nanowires (Nanomaterials.it) with an average length of 10.6 µm and diameter of 81 nm. The metal core of the wires is polycrystalline in nature with a native amorphous oxide coating of 4-8 nm in thickness.

Collection of exponents and prefactors obtained experimentally

The power law scaling behaviour $\Gamma = A\, I_c^{\alpha}$ is ubiquitous across a wide range of nanoscale junction arrangements. Table S1 shows data for the scaling exponent α and the prefactor A for single wires, single junctions and networks made of Ag, Cu, Ni, TiO$_2$ and core shell Ag-TiO$_2$ nanowires. Physical representations of α and A are discussed in the following section.

Table S1: Average α and A values obtained experimentally for a range of nanowire junctions and networks. All values were taken over ensembles of material samples except for the TiO$_2$ single wire from which only one sample was experimented.

| System Type | Material | α | A |
| --- | --- | --- | --- |
| NWN (200 μm) | Ag | 1.10 ± 0.07 | 9.56 ± 10.24 |
| JXN | Ag | 1.05 ± 0.05 | 32.29 ± 15.58 |
| NWN (20 μm) | Cu | 0.97 ± 0.03 | 0.49 ± 0.43 |
| JXN | Cu | 1.05 ± 0.07 | 3.55 ± 2.92 |
| JXN | AgTiO$_2$ | 1.03 ± 0.02 | 7.69 ± 4.77 |
| NWN (50 μm) | Ni | 1.01 ± 0.03 | 0.20 ± 0.21 |
| Single wire | Ni | 0.95 ± 0.01 | 0.29 ± 0.03 |
| Single wire | TiO$_2$ | 0.83 | 0.003 |

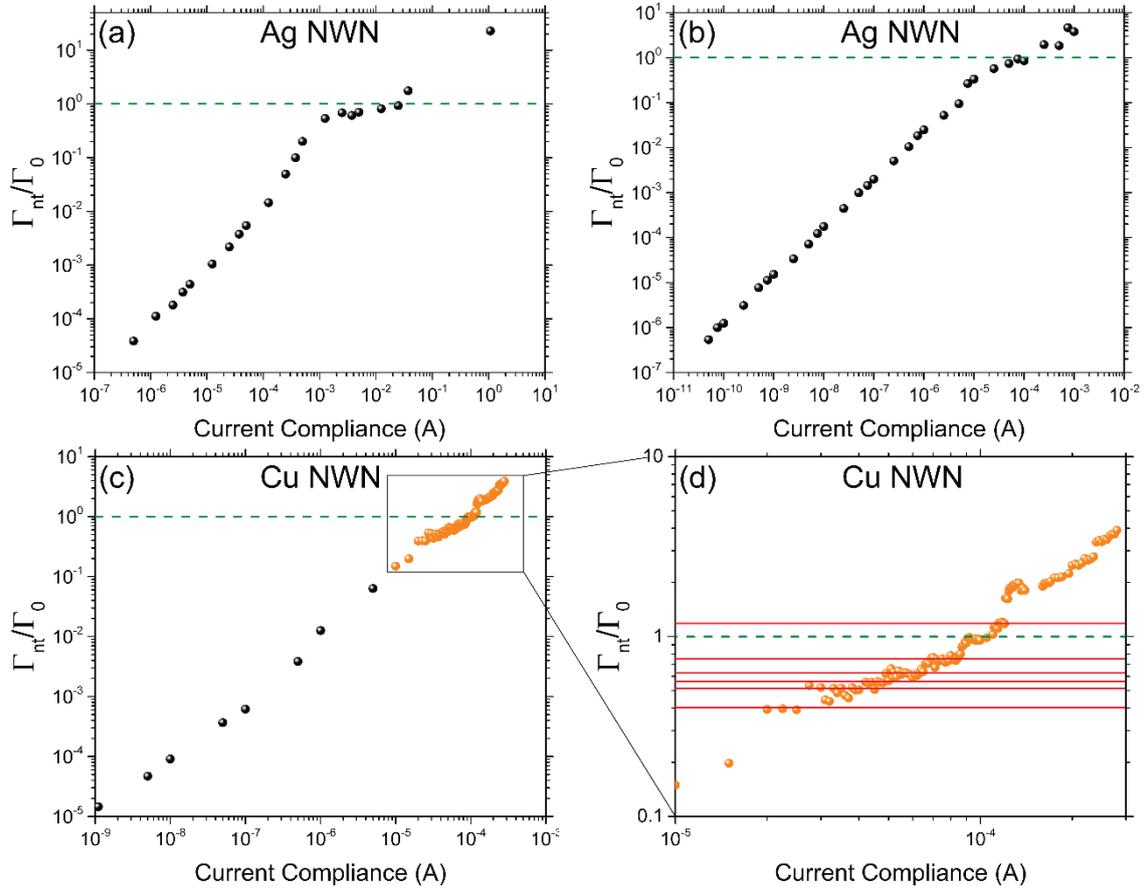

Figure S4: $\Gamma_{nt}/\Gamma_0$ where $\Gamma_{nt}$ is the conductance of the NWN and $\Gamma_0$ is the quantum of conductance plotted against current compliance (log-log scale) with conductance plateaus for Ag NWNs of size (a) 500 x 500 µm and (b) 100 x 100 µm. (c) $\Gamma_{nt}/\Gamma_0$ vs. current compliance values from a 20 x 20 µm Cu NWN. (d) Magnification of the orange data points in panel (c) showing the fine current compliance sampling which detects the numerous conductance plateaus marked by the horizontal red lines. The green dashed line marks where $\Gamma_{nt}/\Gamma_0 = 1$.

## 2. Relationship with the ion-drift model

A popular description of a single memristive system was elaborated by Strukov et al. [7] in which an ion drift mechanism was used to verify the memristor fingerprints of dynamical $TiO_x$-based junctions. Changes in the resistance of the junction are attributed to an effective modulation of the interfacial barrier splitting a doped ($TiO_{2-x}$) and an undoped ($TiO_2$) layer upon application of an electric field. This modulation is caused by the drift of oxygen vacancies across the interface in response to the applied electric field. Following the linear ionic drift in a uniform field assumption, the electrical response of the junction is given by

$$V(t) = \left[R_{on}\frac{w(t)}{D} + R_{off}\left(1 - \frac{w(t)}{D}\right)\right]I(t) \quad (s1)$$

$$\frac{dw}{dt} = \mu_v \frac{R_{on}}{D} I(t) \quad (s2)$$

where t is the time, D is the full length of the junction, μ$_v$ is the ion mobility, I is the input current, and V is the output potential (considering that the system is current-driven). w is a state variable representing the varying length of the doped layer.

We will now demonstrate how the state equation (s2) can be obtained from the power-law relation in equation (1) (main text) considering another proxy for the conductance scaling. According to our PL+C description, the conductance of a single junction is a dynamical quantity whose value is regulated by the current compliance. For each I$_c$ value, a current versus time curve is unfolded out of a full voltage sweep; its area is given by $Q_c = \int_{-\infty}^{t} I(\tau)d\tau$ where Q$_c$ is the total amount of charge flowing through the junction during the time period of a full voltage sweep given by t. Therefore, an increment in I$_c$ yields an increment in Q$_c$ in such a way that $Q_c \propto I_c$. Without loss of generality, equation (1) (main text) can be then written in terms of the cumulative charge Q$_c$ such as,

$$\Gamma_j = A_j Q_c^{\alpha_j} \quad (s3)$$

with the Q$_c$ being the new proxy variable of the PL. In fact, all our experimental results taken for junctions and networks were minimally affected with respect to plotting the conductance data as a function of Q$_c$ or I$_c$.

The junctions studied in this work spend most of their lifespan "stabilized" in the non-resonant tunnelling regime in which their conductance follows an exponential dependency with the electrode separation,

$$\Gamma_j(t) = \Gamma_0 e^{-\beta[D-w(t)]} \quad (s4)$$

where β is the decay parameter that characterizes the tunnelling barrier and w(t) here represents the length of the conducting filament. For sufficiently small tunnelling separations and considering the simplest case in which $\alpha_j = 1$, we can approximate equation (s4) up to first order and related it with equation (s3) such as

$$\Gamma_j(t) \approx \Gamma_0[1 - \beta(D - w(t))] = A_j Q_c(t) \quad (s5)$$

Performing the derivative in time of equation (s5) we obtain

$$\frac{d\Gamma_j}{dt} = \Gamma_0 \beta \frac{dw}{dt} = A_j \frac{dQ_c}{dt} \qquad (s6)$$

which gives the following state equation ruling the filament growth

$$\frac{dw}{dt} = \frac{A_j}{\beta \Gamma_0} I(t) \qquad (s7).$$

Considering that $R_{on} = 1/\Gamma_0$, the comparison between equations (s2) and (s7) provides

$$A_j = \frac{\mu_v \beta}{D} \qquad (s8).$$

The effects of nonlinearity in the charge carrier drift can be captured by repeating the whole derivation above for $\alpha_j \neq 1$. Using equations (s3) and (s4), we get the generalized state equation

$$\frac{dw}{dt} = \frac{\mu_v}{D\Gamma_0} I(t)\, e^{\beta[D-w(t)]}\, \alpha_j [Q_c(t)]^{\alpha_j - 1} = \frac{\mu_v}{D\Gamma_0} I(t)\, f\left(w/D, \alpha_j, Q_c(t)\right) \qquad (s9).$$

Note that the first order approximation in equation (s4) was not taken in this derivation rendering hence the exponential term in expression (s9). This result is analogue to other generalizations of the ion-drift model [7, 8, 9, 10, 11, 12] in which window functions are used to account for the nonlinear effects existent at the boundaries of the conducting channel and for the nonlinear dependency of the state derivative on the driven current.

## 3. Graphic representation of PL+C model for individual junctions

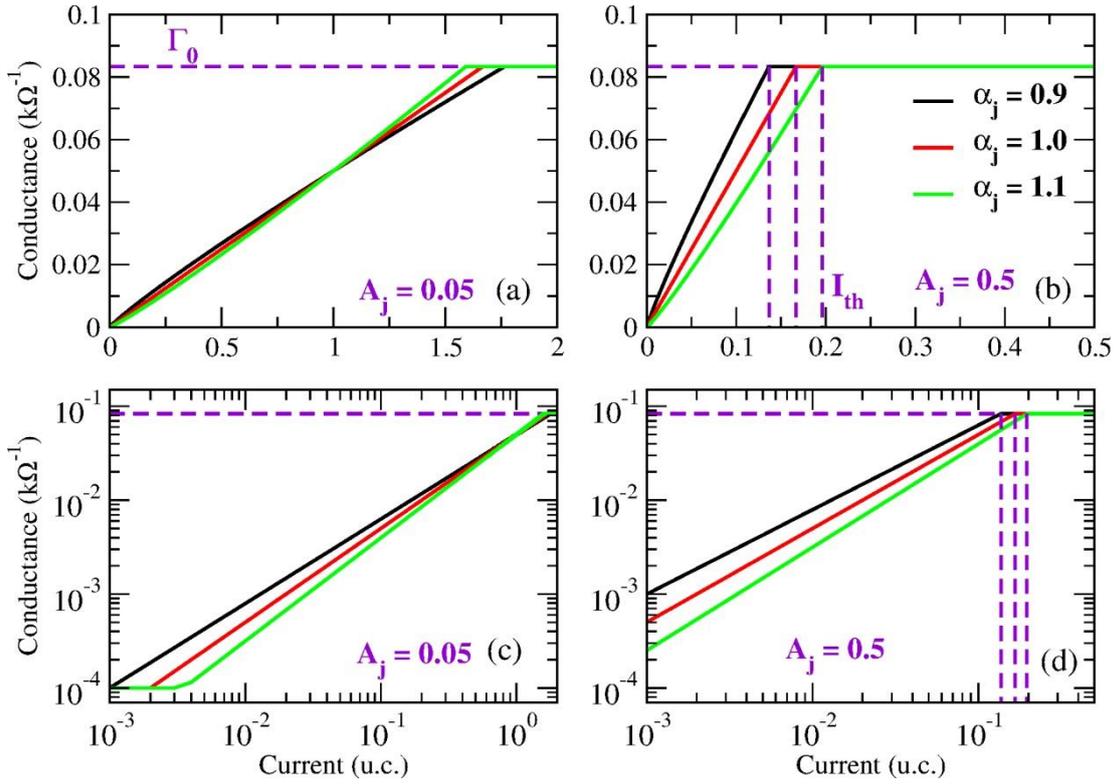

Figure S5: Conductance versus current plots taken for a single junction at linear (upper panels) and logarithmic (lower panels) scales. The current is expressed in units of current (u.c.). The curves were taken using PL+C model as specified in the main text. The prefactor value used in the left panels [(a) and (c)] was $A_j = 0.05$ and in the right panels [(b) and (d)] was $A_j = 0.5$. Horizontal dashed lines mark the quantum of conductance whereas vertical ones mark the current thresholds up to where a junction displays memristive features. This is given by $I_{th} = (1/A_j R_{on})^{1/\alpha_j}$.

# 4. "Part-to-whole" computational algorithm for NWNs and network templates

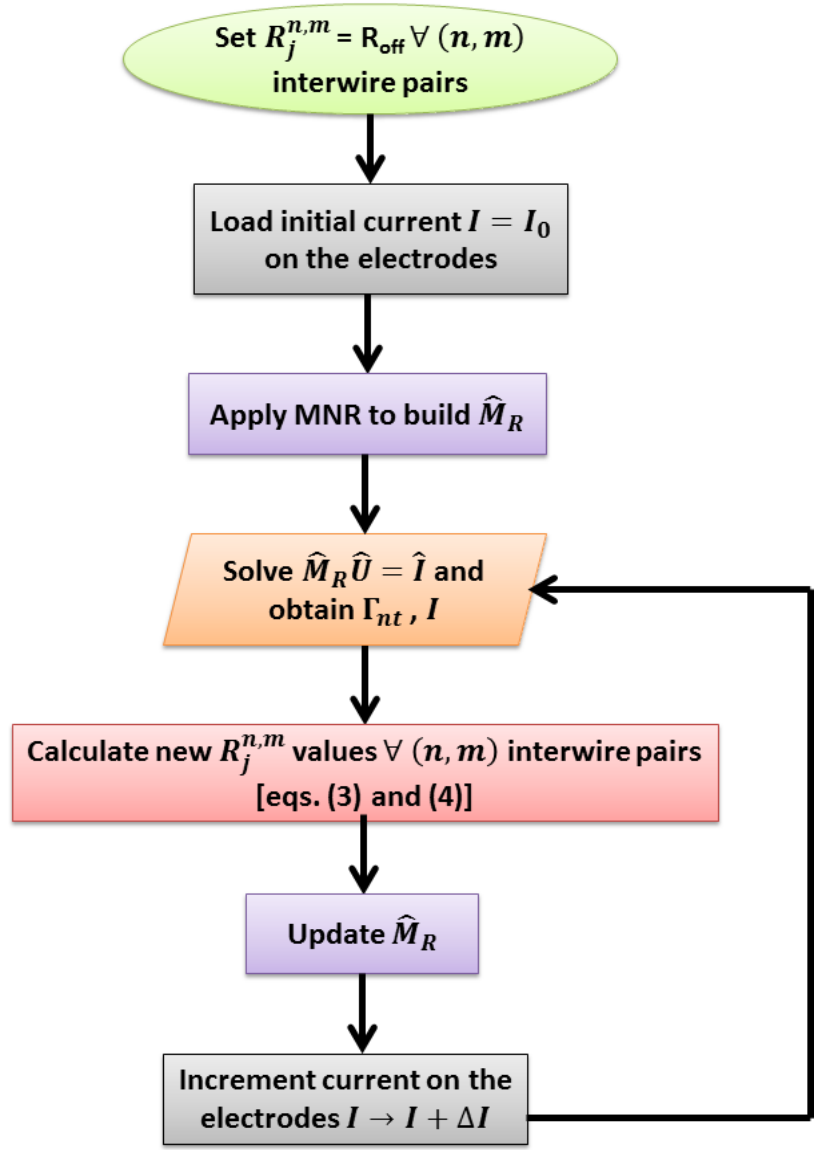

Figure S6: Schematics of the computational implementation of PL+C junction model onto macroscopic networks. Equations (3) and (4) specified in this workflow refers to equations (3) and (4) appearing in the main text. The algorithm obtains the conductance evolution of NWNs subjected to an electrical current source. See main text for detailed explanation of the algorithm.

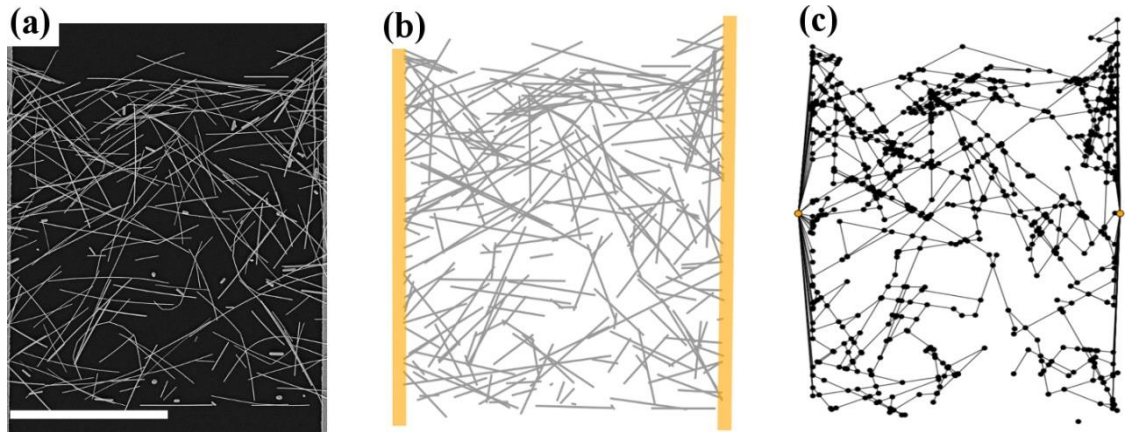

Figure S7: (a) Scanning electron micrograph (SEM) of a silver nanowire network containing 0.49 NWs/µm$^2$. Bottom scale bar represents 10 µm. (b) Computational transcription of the network after image processing the image in (a). Wires are represented by grey sticks and electrodes are represented by vertical orange bars. (c) Graph representation of the virtual NWN that serves as a template for the current maps shown in Figures 3 (main text), S9, and S10.

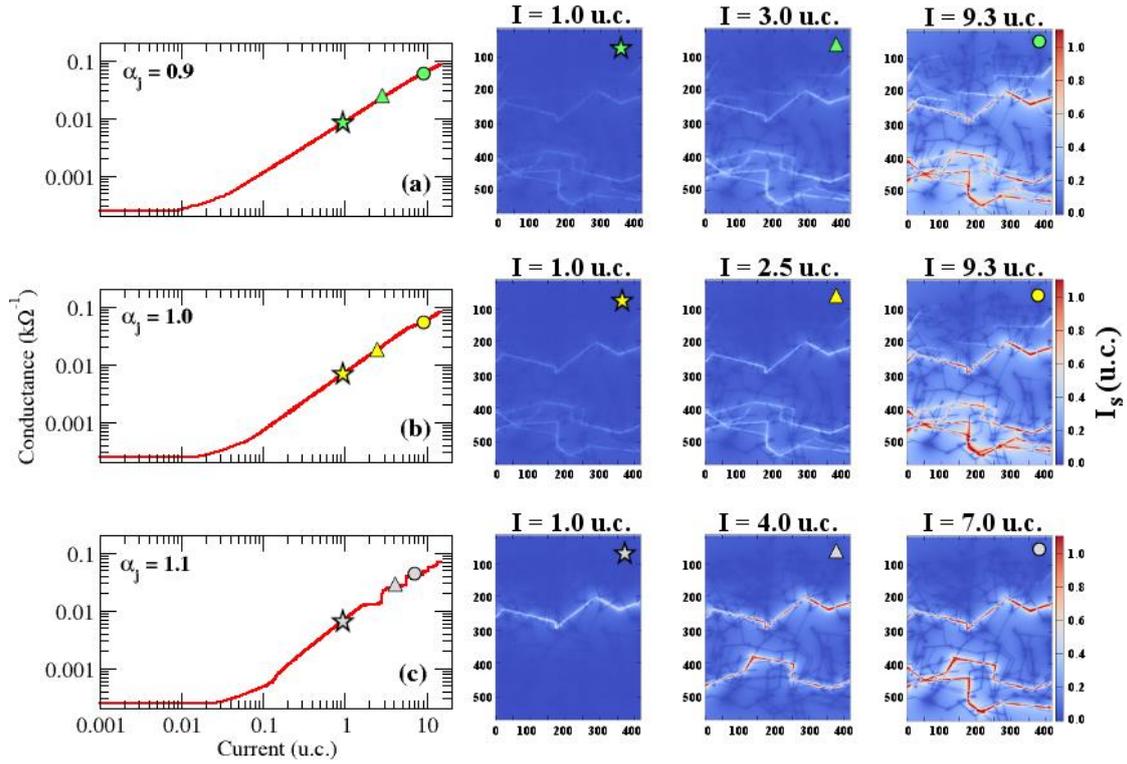

Figure S8: (Left panels) Conductance versus current curves for an Ag NWN sample containing 0.47 NWs/μm². The junctions were described with $A_j = 0.05$ and exponents of (a) $\alpha_j = 0.9$, (b) $\alpha_j = 1.0$, (c) $\alpha_j = 1.1$. The symbols mark points in the curves in which current colour maps were taken. (Right panels) Current colour maps calculated over each wire segment ($I_s$) of the Ag NWN. Snapshots were taken for three current values specified on the top of each current map and distinguished by the symbols: star (set in the PL regime), and triangle and circle (both set in the PPL regime). Animations revealing the complete evolution of the network in response to the current source, junction optimization of the top-3 paths of least-resistance, and current-segment maps are provided in the supplemental material (cf. Section 7 of this document for animation description).

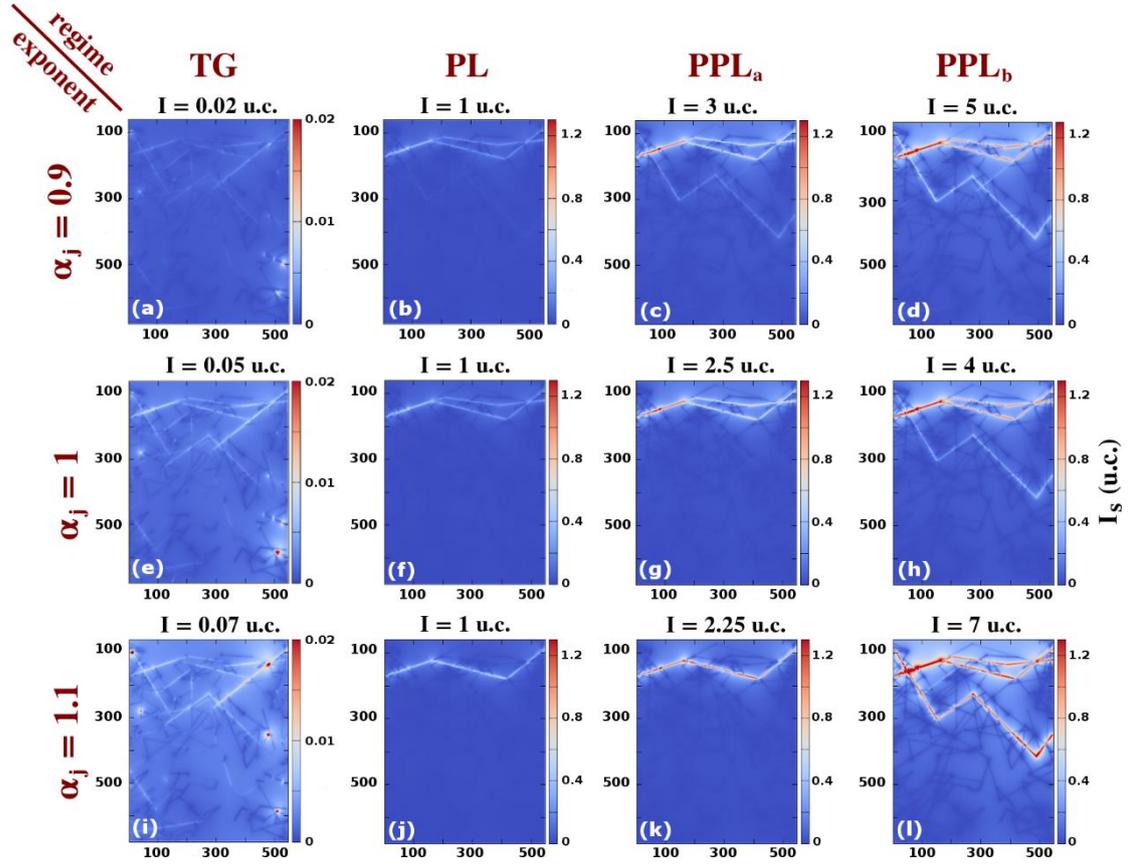

Figure S9: Current colour maps calculated over each wire segment ($I_s$) of the Ag NWN in Figure S7. Snapshots are arranged in a grid in which its abscissa refers to the conducting regime and its ordinate refers to the junction exponent. These maps were obtained for a fixed prefactor value of $A_j = 0.05$ and for distinct junction exponents: (top) $\alpha_j = 0.9$, (middle) $\alpha_j = 1.0$, (bottom) $\alpha_j = 1.1$. The current values in which the snapshots were taken are located at the TG, PL, and two PPLs (labelled as <u>a</u> and <u>b</u>) regimes. Bottom panels correspond to the same results shown in Figure 3 in the main text. Animations revealing the complete evolution of the network in response to the current source, junction optimization of the top-3 paths of least-resistance, and current-segment maps are provided in the supplemental material (cf. Section 7 of this document for animation description).

## 5. Current maps for a NWN with disorder in the junction exponents and quick activation element

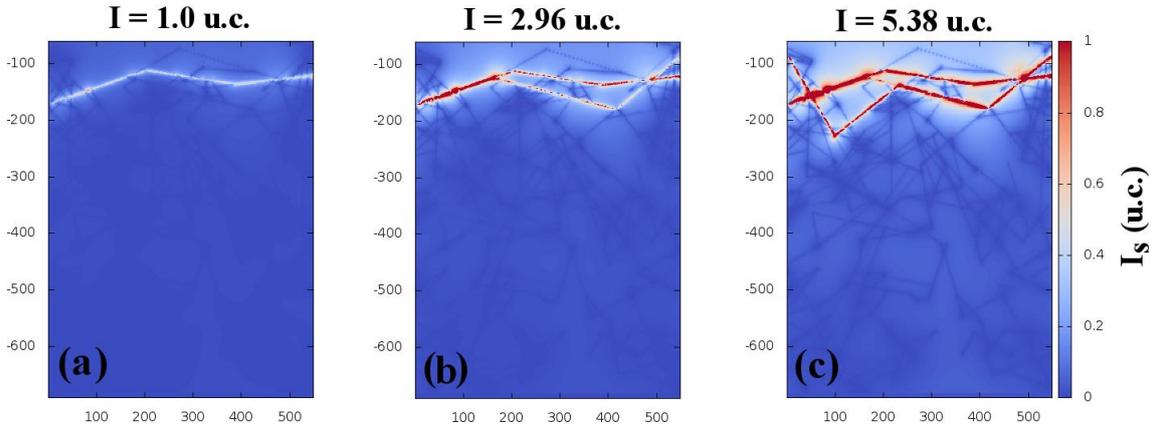

Figure S10: Current colour maps calculated over each wire segment ($I_s$) of the Ag NWN depicted in Figure S7. The junction settings are for a heterogeneous NWN with $A_j = 0.05$ and a narrow dispersion was induced in the $\alpha_j$ exponents using a normal distribution with $\langle \alpha_j \rangle = 1.05$, $\sigma = 0.1$ and truncated at [1.0,1.1]. Snapshots were taken for three input currents corresponding to different conducting regimes: (a) I = 1 u.c. (PL), (b) I = 2.96 u.c. (PPL), and (c) I = 5.38 u.c. (PPL).

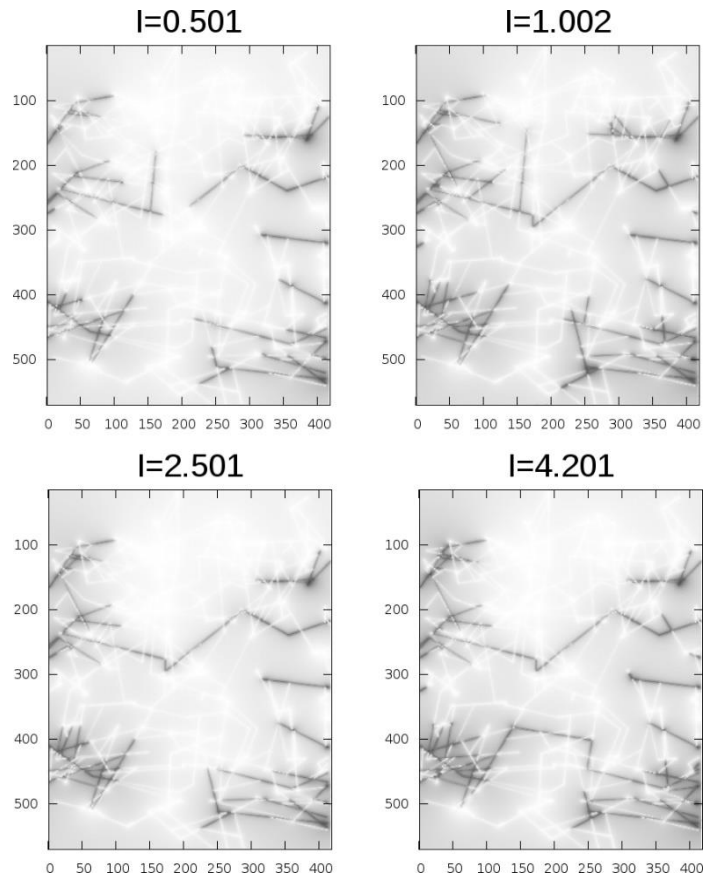

Figure S11: Binary activation maps taken for the Ag NWN investigated in Figure S8. Four snapshots were obtained for distinct input current values written in units of current (u.c.): $I \cong$

0.5, 1.0, 2.5, 4.2 u.c.. Only wires carrying an amount of current above a certain threshold value (>0.009 u.c.) were contrasted in dark colour. This result was taken with the assumption that NWs in the vicinity of the source electrode activate faster than the ones located in the mid-section of the network. Note the more imperfect profile of the WTA state - which emerges at $I \cong 2.5$ u.c. - with numerous wires branching out of the electrodes as a result of current leakage to other parts of the network. Yet, the WTA state manages to carry about 81% of the total current as quantified within our simulations.

## 6. Influence of NWN inner resistance on the self-similar behaviour

Let's assume the simplified case in which the network selects <u>M</u> identical parallel paths, each path containing the same number of wires <u>N</u>. Therefore, each path contains N segments and N+1 junctions (including the two extra junctions that the wire path makes with the electrodes). Let's also consider that all wire segments have nearly the same length. The resistance of a single path $(R_p)$ can be then written as

$$R_p = NR_{in} + (N+1)R_j \qquad (s10).$$

Substituting $R_j = i^{-\alpha_j}/A_j$ (junctions ruled by a power law) into (s10), we obtain

$$R_p = NR_{in} + (N+1)\frac{i^{-\alpha_j}}{A_j} \qquad (s11)$$

being $i$ the amount of current flowing through the path, i.e. $i = I/M$ with $I$ being the current sourced in the electrodes. We can then write down the equivalent conductance for <u>M</u> parallel paths as

$$\Gamma_{eq} = \Gamma_{nt} = \frac{M}{R_p} \qquad (s12).$$

Further manipulation of expression (s12) gives

$$\Gamma_{nt} = \left(\frac{MA_j}{N+1}\right)i^{\alpha_j}\left[\frac{1}{1 + \left(\frac{NR_{in}A_j}{N+1}\right)i^{\alpha_j}}\right] \qquad (s13)$$

which shows that the dominant contribution for the network conductance in the limit of low inner wire resistances ($R_{in} \ll R_j$) is a power-law with $\alpha_{nt} = \alpha_j$ and $A_{nt} = M^{1-\alpha_j}A_j/(N+1)$. The term inside the square brackets in equation (s13) will play a relevant role when the skeleton of the network becomes more robust, i.e. $R_{in}$ increases. In this case one expects a breakdown in the self-similar behaviour as the functional form of $\Gamma_{nt}(I)$ does not correspond to a pure power law. To confirm this point, we conduct extra calculations on the same network of Figure S7 except that the resistivity of the wires was artificially increased as $\rho = 100\rho_{Ag}$. Figure S12 shows the comparison between the real ($\rho = \rho_{Ag}$) and the hypothetical network ($\rho = 100\rho_{Ag}$) with junctions described by $\{A_j = 0.05, \alpha_j = 1.1\}$. One can see that the conductance of the network made with highly resistive wires shows a pronounced deviation from the pure power law $\Gamma_{nt} \propto I^{1.1}$, confirming the premise raised by equation (s13).

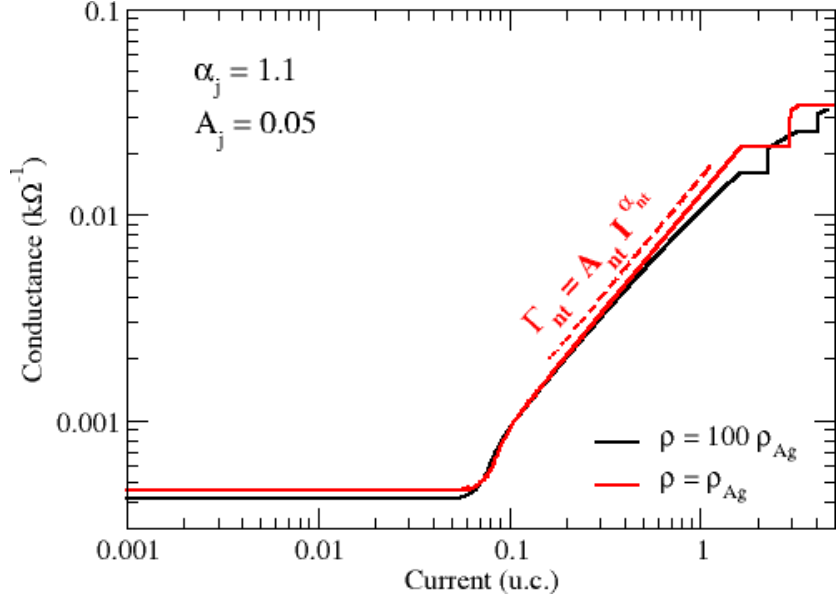

Figure S12: Conductance versus current curves taken for the Ag NWN shown in Figure S7 with wire resistivity of $\rho = \rho_{Ag} = 22.6 \pm 2.3 \, n\Omega m$ (red line) and $\rho = 100\rho_{Ag}$ (black line). The dashed red line shows a power law fitting of $\Gamma_{nt} = A_{nt}I^{\alpha_{nt}}$ with $A_{nt} = 0.0125$ and $\alpha_{nt} = 1.1$. Currents are expressed in units of current (u.c.). The junctions were described using $\{A_j = 0.05, \alpha_j = 1.1\}$.

## 7. Description of the animation supporting files

Animations revealing the complete evolution of the network in response to the current source are provided in numerous supporting files. Each snapshot contains six panels as explained in the sketch of Figure S13.

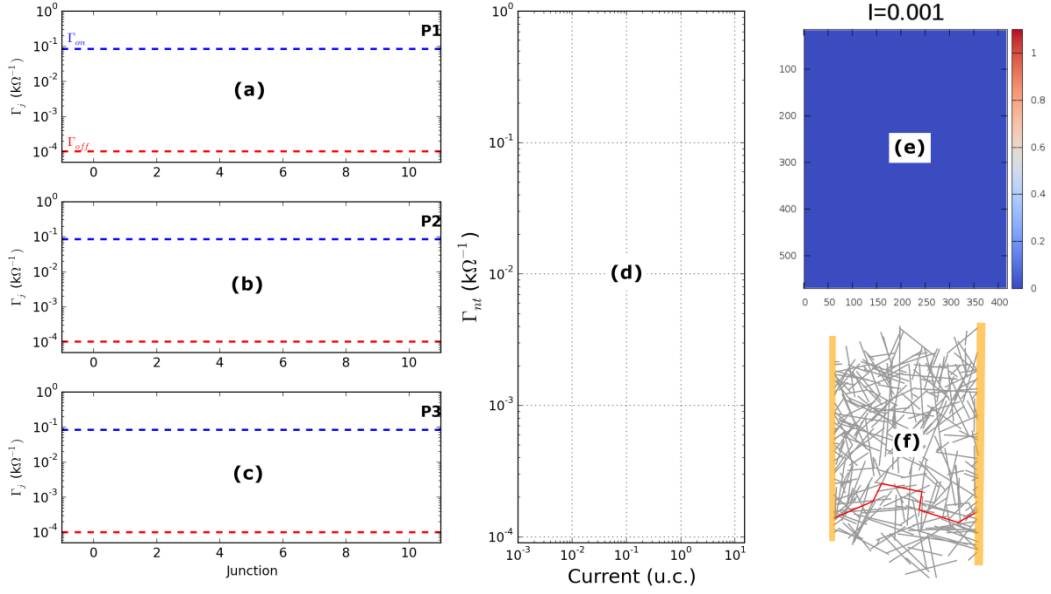

Figure S13: Snapshot template of the animation revealing the conductance evolution of a NWN as a function of current. Panels (a-c) monitor the conductance of the junctions $(\Gamma_j)$ that compose the top-3 paths carrying most of the current load. Junctions are labelled as integer numbers in the abscissa. The conductance values are displayed as level-bars that are free to move in the range of $[\Gamma_{off}, \Gamma_{on}]$ marked by horizontal dashed lines. Panel (d) shows the whole $\Gamma_{nt} \times I$ curve with a dynamical marker (red circle) that tracks the evolution of the simulation. (e) Current-segment maps as those depicted in Figures 3 (main text), S8, S9, and S10. (f) Digitalized image of the NWN with the top-3 paths being highlighted with red sticks.

The names of the animation files and their respective initial conditions are described below:

- ani_n049_Aj005_exp_1.m4v: animation for the Ag NWN shown in Figure S7 with $\{A_j = 0.05, \alpha_j = 1\}$. One can see the formation of two superimposed conductive paths in the first power-law regime. Subsequent paths are formed as current increases causing the observed changes in slopes in the conductance curve.
- ani_n049_Aj005_exp_11.m4v: animation for the Ag NWN shown in Figure S7 with $\{A_j = 0.05, \alpha_j = 1.1\}$. One can see the formation of a single conductive path in the power-law regime. Once all the junctions in this path are fully optimized, the network becomes temporarily Ohmic, i.e. its conductance does not change within a certain current window. Further paths are formed in a quantized manner as current is loaded onto the electrodes with the conductance curve depicting a stepwise increase.
- ani_n047_Aj005_exp_1.m4v: animation for the Ag NWN used in Figure S8 with

$\{A_j = 0.05, \alpha_j = 1\}$. One can see the formation of multiple conductive paths in the first power-law regime. Subsequent paths are formed as current increases causing the observed changes in slopes in the conductance curve.

- ani_n047_Aj005_exp_11.m4v: animation for the Ag NWN used in Figure S8 with $\{A_j = 0.05, \alpha_j = 1.1\}$. One can see the formation of a single conductive path practically slicing the network at half in the power-law regime. Once all the junctions in this path are fully optimized, the network becomes temporarily Ohmic, i.e. its conductance does not change within a certain current window. After the first conductance plateau, one can observe the formation of two independent conductive paths. As more current is loaded onto the terminals, additional paths are formed in a quantized manner with the conductance curve depicting a stepwise increase.

## 8. PVC SEM images of large NWNs

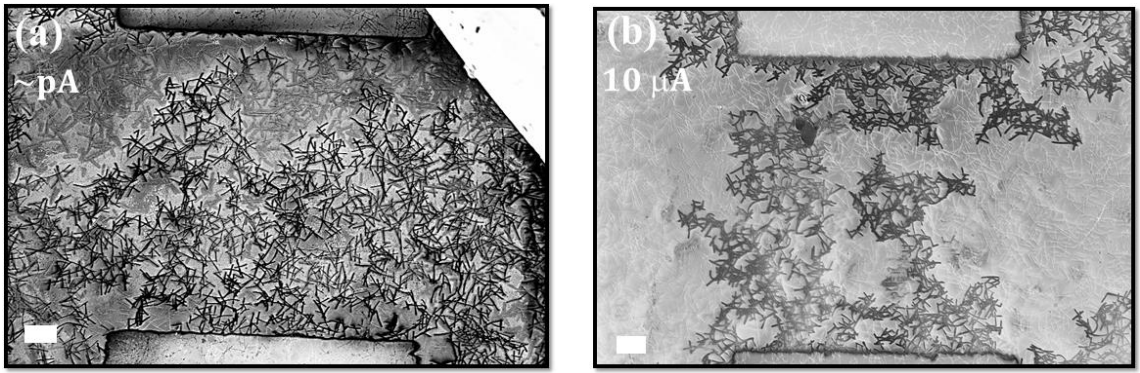

Figure S14: PVC SEM images of two distinct Ag NWN samples of dimensions 200 x 200 µm. In (a), the image was taken by holding the source voltage of 2 V resulting in a leakage current of a few hundreds of pA (see label on image). On panel (b), the image was taken during an I-V sweep (electrodes grounded) with limiting current compliances of 10 µA. Current levels are written on the respective panels. White scale bars correspond to 20 µm. Note a PVC image provides a qualitative comparison of the electrical connectivity within a given network. It is not possible to compare the contrast observed in different networks or even that observed in the same network imaged under different conditions.

**REFERENCES:**

[1] Valov I, Kozicki MN. Cation-based resistance change memory. *Journal of Physics D: Applied Physics* **46**, 074005 (2013).

[2] Jo SH, Lu W. CMOS compatible nanoscale nonvolatile resistance switching memory. *Nano Letters* **8**, 392-397 (2008).

[3] Schrögmeier P, et al. Time discrete voltage sensing and iterative programming control for a $4F^2$ multilevel CBRAM. *2007 IEEE Symposium on VLSI Circuits Digest of Technical Papers*, DOI: 10.1109/VLSIC.2007.4342708.

[4] Gopalan C, et al. Demonstration of conductive bridging random access memory (CBRAM) in logic CMOS process. *Solid-State Electronics* **58**, 54-61 (2011).

[5] Du J, Zhang J, Liu Z, Han B, Jiang T, Huang Y. Controlled synthesis of $Ag/TiO_2$ core-shell nanowires with smooth and bristled surfaces via a one-step solution route. *Langmuir* **22**, 1307-1312 (2006).

[6] Rathmell AR, Wiley BJ. The synthesis and coating of long, thin copper nanowires to make flexible, transparent conducting films on plastic substrates. *Advanced Materials* **23**, 4798-4803 (2011).

[7] Strukov DB, Snider GS, Stewart DR, Williams RS. The missing memristor found. *Nature* **453**, 80-83 (2008).

[8] Wey T, Benderli S. Amplitude modulator circuit featuring $TiO_2$ memristor with linear dopant drift. *Electronics Letters* **45**, 1103-1104 (2009).

[9] Joglekar YN, Wolf SJ. The elusive memristor: properties of basic electrical circuits. *European Journal of Physics* **30**, 661 (2009).

[10] Biolek Z, Biolek D, Biolkova V. SPICE model of memristor with nonlinear dopant drift. *Radioengineering* **18**, 210-214 (2009).

[11] Prodromakis T, Peh BP, Papavassiliou C, Toumazou C. A versatile memristor model with nonlinear dopant kinetics. *IEEE Transactions on Electron Devices* **58**, 3099-3105 (2011).

[12] Lehtonen E, Laiho M. CNN using memristors for neighborhood connections. *2010 12th International Workshop on Cellular Nanoscale Networks and their Applications (CNNA)*, DOI: 10.1109/CNNA.2010.5430304.